\documentclass[prl,twocolumn,preprintnumbers,amsmath,amssymb]{revtex4}

\usepackage{graphicx}
\usepackage[ansinew]{inputenc}
\usepackage[T1]{fontenc}
\usepackage{color}
\usepackage{ae,aecompl}

\begin{document}

\title{Shaping the Phase of a Single Photon}

\author{H.~P.~Specht}
\author{J.~Bochmann}
\author{M.~M{\"{u}}cke}
\author{B.~Weber}
\author{E.~Figueroa}
\author{D.~L.~Moehring}
 \email{david.moehring@mpq.mpg.de}
\author{G.~Rempe}

\affiliation{Max-Planck-Institut für Quantenoptik, Hans-Kopfermann-Strasse~1, 85748 Garching, Germany}

\maketitle

\textbf{
While the phase of a coherent light field can be precisely known, the phase of the individual photons that create this field, considered individually, cannot \cite{carruthers:1968}.  Phase changes \textit{within} single-photon wave packets, however, have observable effects.  In fact, actively controlling the phase of individual photons has been identified as a powerful resource for quantum communication protocols \cite{tittel:2000, inoue:2002}.  Here we demonstrate the arbitrary phase control of a single photon.  The phase modulation is applied without affecting the photon's amplitude profile and is verified via a two-photon quantum interference measurement \cite{hong:1987, shih:1988}, which can result in the fermionic spatial behaviour of photon pairs.  Combined with previously demonstrated control of a single photon's amplitude \cite{kuhn:2002, keller:2004, mckeever:2004, bochmann:2008, kolchin:2008}, frequency \cite{legero:2004}, and polarisation \cite{wilk:2007}, the fully deterministic phase shaping presented here allows for the complete control of single-photon wave packets.  
}

\begin{figure}[htb]
\includegraphics[width=1.0\columnwidth,keepaspectratio]{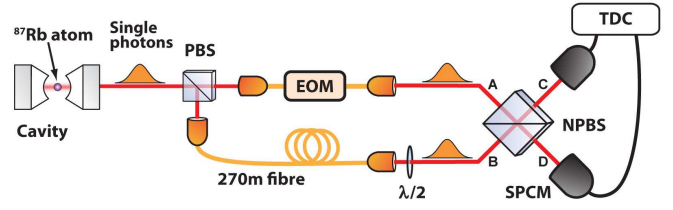}
\caption{
\textbf{Experimental apparatus.}
A single $^{87}$Rb atom is coupled to the optical mode of a high-finesse cavity, which when subjected to a sequence of laser pulses, emits a stream of single photons.  A polarising beam splitter (PBS) directs the photons randomly into one of two optical paths.  In the first path, an electro-optic modulator (EOM) is used to arbitrarily shape the phase of the single photon wave packets.  A delay fibre in the second optical path allows two subsequently generated photons to synchronously impinge on a 50/50 non-polarising beam splitter (NPBS).  The photons are detected by single photon counting modules (SPCMs) which are connected to a time to digital converter (TDC).
}
\label{fig:setup}
\end{figure}

Consider two identical photons mode-matched at the two input ports ($A$ and $B$) of a 50/50 non-polarising beam splitter (NPBS), represented by the initial state $|\Psi_{i}\rangle=|1_{A}1_B\rangle$ (see Fig.~\ref{fig:setup}).  Due to the indistinguishability of the photons, the detection of one photon in output port $C$ or $D$ at time $t_{0}$ projects the initial product state $|\Psi_{i}\rangle$ into the ``which path'' superposition state $|\Psi_{\pm}(t_{0})\rangle= (|1_{A}, 0_{B}\rangle\pm|0_{A}, 1_{B}\rangle)/\sqrt{2}$ of the remaining photon.  As first demonstrated by Hong, Ou and Mandel \cite{hong:1987}, the bosonic nature of photons always results in the detection of the second photon in the same output port as the first.  However, we can alter this coalescence behaviour by introducing an arbitrary differential phase $\Delta_\phi$ between the two components of $|\Psi_{\pm}\rangle$.  This results in a phase-dependent wave function of the remaining single photon
\begin{equation}
|\tilde{\Psi}_{\pm}(t_0+\tau)\rangle = (|1_{A},0_B\rangle\pm e^{i\Delta_\phi(\tau)}|0_{A},1_B\rangle)/\sqrt{2}.
\label{eq:interference}
\end{equation}
From (\ref{eq:interference}), it can be readily shown that the probability to detect the two photons in opposite output ports depends on the magnitude of the applied phase shift \textit{between} the detection times \cite{legero:2004, legero:2006}:
\begin{equation}
p_{\text{coinc}}=\text{sin}^2(\Delta_\phi/2).
\label{eq:TheFormula}
\end{equation} 
Hence, with well tailored phase shifts, we gain control over the statistics of coincident photon detections.

In our experiment, single photons are generated from a coupled single-atom cavity QED system (see Fig.~\ref{fig:setup} and the Methods section).  This apparatus allows us to produce photons at the output mode of the cavity with a well defined spatio-temporal mode profile \cite{kuhn:2002, keller:2004, mckeever:2004, bochmann:2008}.  In particular, we design the photons with a Gaussian temporal profile --- a shape proven to be optimal for two photons interacting on a beam splitter \cite{rohde:2005}.  Our photon wave packets have a full width at half maximum of 150~ns, much longer than our single photon detection time resolution of 2~ns.  The single-photon character of our photon stream is verified by measuring the second-order correlation function $g^{(2)}(\tau)$.  The resulting suppression of coincidence events of $95\pm1\%$ is consistent with independently measured influences of multi-atom events within the cavity and noise on the single photon counting modules (SPCMs).

The single photons are generated with a random polarisation such that a polarisation beam splitter (PBS) equally distributes them into two optical paths (Fig.~\ref{fig:setup}).  In one path, a fibre based electro-optic phase modulator (EOM) is used to arbitrarily shape the phase of the photonic wave packets.  The other optical path consists of a 270~m long delay fibre.  These two optical paths are finally mode-matched on a NPBS, after which the photons are detected by SPCMs connected to a time-to-digital converter.  The photons are generated at a rate of 740~kHz, chosen to match the path length difference between the EOM fibre and the delay fibre.  With this configuration, two independent, subsequently generated photons can arrive at the beam splitter simultaneously \cite{santori:2002, legero:2004, wilk:2007}.    

\begin{figure}[htb]
\includegraphics[width=1.0\columnwidth,keepaspectratio]{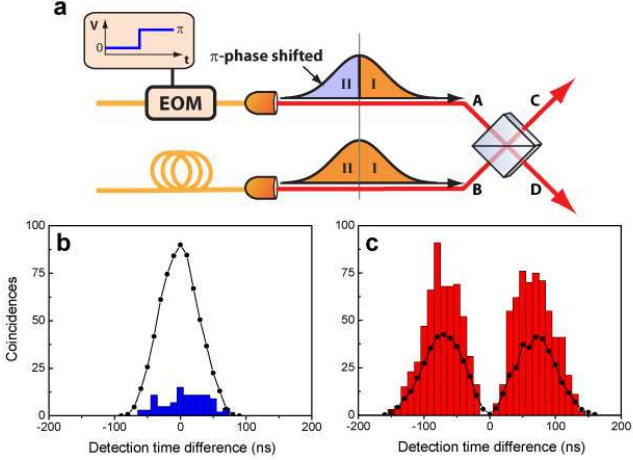}
\caption{
\textbf{Photon pair coincidences.}
\textbf{a}, One of the two photons of each pair is subjected to a $\pi$-phase shift at the centre of the wave packet envelope via a voltage step applied to an EOM.
\textbf{b,c}, Connected dots refer to measured coincidences for non-interfering photons (see the Methods section) and histograms refer to measured coincidences of interfering photons.  
\textbf{b}, Interfering photons detected in the same temporal half of the photon wave packets (both within I or both within II) should never be detected in opposite output ports.  The total measured coincidence detection rate is only $16\pm2\%$ of that for non-interfering photons.
\textbf{c}, Interfering photons detected in opposite halves of the wave packet (one within I and the other within II) are ideally twice as likely to be detected in opposite output ports as compared to non-interfering photons (i.e. always).  The measured coincidence rate is $183\pm6\%$ that for non-interfering photons.  All stated uncertainties are statistical.
Due to the finite rise time of the phase shift, photons detected within $\pm5$~ns of the step are omitted from the above data.  Photons detected further than 80~ns away from the step are also omitted due to the low signal to noise ratio.
}
\label{fig:pi-phase}
\end{figure}

In a first experiment, we demonstrate the ability to convert the usual bosonic coalescence for photon pairs into anti-coalescence by introducing a relative phase shift of $\Delta_{\phi}=\pi$ between the photon detections.  This is accomplished by subjecting the photon traveling through the EOM to a sudden $\pi$ phase shift exactly at the centre of the wave packet envelope (Fig.~\ref{fig:pi-phase}).  This results in the total wave functions of the two photons to be orthogonal to each other.  Hence, when integrating over the entire wave packets, it is equally likely to detect the two photons in the same or opposite detectors.  However, closer examination reveals that the photon counting statistics depend on when the photons are detected with respect to the applied phase shift, as only a phase shift between photon detections is relevant.  For instance, if both photons are detected within the same temporal half of the wave packet envelopes (as defined in Fig.~\ref{fig:pi-phase}a), then there is no relative applied phase shift ($\Delta_\phi=0$), and the photons coalesce.  On the other hand, if the photons are detected in opposite temporal halves ($\Delta_\phi=\pi$), then according to equation (\ref{eq:TheFormula}), the photons are always detected in opposite output ports.  These coincidences are therefore twice as likely as for non-interfering photons, equivalent to a fermionic spatial behaviour. 

\begin{figure}
\includegraphics[width=1.0\columnwidth,keepaspectratio]{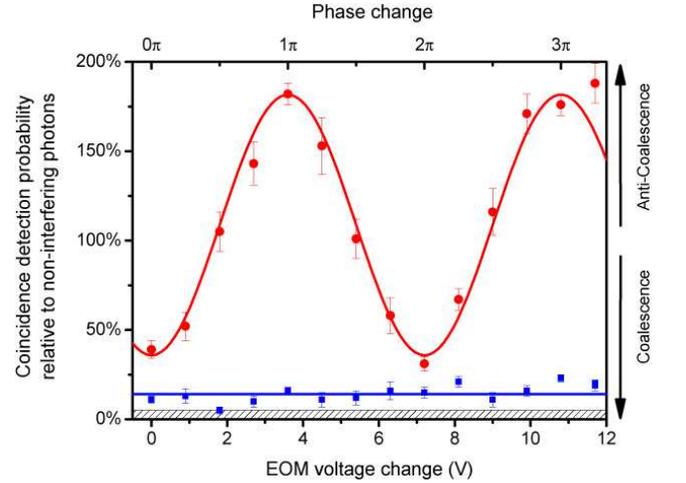}
\caption{
\textbf{Coincidence rate vs. applied phase shift.}
Red dots show the coincidence detection probability with respect to non-interfering photons for photon pairs detected in opposite temporal halves of the photon wave packet.  The solid fitted line follows the sinusoidal behaviour of equation (\ref{eq:TheFormula}) with a visibility of $67\pm3\%$.  The coincidence rate for photon pairs detected in the same half of the photon wave packet remains at a low value of $\approx14\%$ (individual points: blue squares; average value: solid blue line).  An offset of 5\% due solely to multi-atom effects and noise on the SPCMs is indicated by the filled area.  The phase shift of the EOM is proportional to the applied voltage, verfied via an independent measurement with continuous laser light, with a resulting $\pi$-voltage of $3.60\pm0.05V$.  The data from Figure~\ref{fig:pi-phase}b-c is incorporated into this Figure.
}
\label{fig:oscillations}
\end{figure}

The results of our experiment are shown in Figure~\ref{fig:pi-phase}b-c.  When both photons are detected in the same temporal half of their wave packets, they exhibit a strong coalescence resulting in a coincidence rate only $16\pm2\%$ of that for non-interfering photons (see the Methods section).  On the contrary, photons detected in opposite temporal halves show anti-coalescence, where the detection of photon pairs in different output ports is $183\pm6\%$ as likely as for non-interfering photons (Fig.~\ref{fig:pi-phase}c).  The deviation from the respective ideal values of $0\%$ and $200\%$ can be explained by our non-perfect single photon source as well as by mode mismatch on the beam splitter.  In the latter case, the spatial and polarisation modes overlap with greater than 99\% fidelity.  However, shot-to-shot variations in the amplitude and frequency profile of the emitted photons, due largely to addressing different Zeeman sub-levels in the photon generation process, can lead to a temporal or frequency mismatch on the beam splitter \cite{legero:2004, legero:2006}. 

As mentioned previously, when integrated over the entire wave packets (sum of the data from Figs.~\ref{fig:pi-phase}b and c), the interference effects vanish and the total detected coincidences are $98\pm3\%$ compared to non-interfering photons.  These results are analogous to the case of orthogonally polarised photons --- while a vertically and a horizontally polarised photon are non-interfering on a NPBS, perfect coalescence (anti-coalescence) will still arise when the photons are detected with parallel (orthogonal) polarisations in diagonal bases \cite{kwiat:1992}.  

Next, we extend our experiment to investigate the effect of discrete phase shifts with arbitrary magnitude $\Delta_\phi$.  As seen in Figure~\ref{fig:oscillations}, photons detected in the same temporal half of the wave packet envelopes always coalesce, independent of the applied phase shift.  However, photons detected in opposite temporal halves exhibit a $\text{sin}^2(\Delta_\phi/2)$ dependence of the photon pair coincidence detection rate, as given by equation (\ref{eq:TheFormula}), with a visibility of $67\pm3\%$.  As expected, a maximum number of coincidences is observed for phase shifts of odd multiples of $\pi$, whereas the photons coalesce for even multiples of $\pi$.  The fact that our visibility remains consistent even for high applied phase shifts testifies to the stability of our phase modulation, determined mainly by the voltage generator.  In the case of $\Delta_\phi=0,2\pi$, the coincidence rates for photons detected in opposite temporal halves of the photon wave packets should be the same as for those detected in the same temporal half.  However, we observe a small discrepancy due to accumulated phase noise from the frequency and amplitude uncertainty in our single photon stream.  This leads to an increased coincident detection rate for higher detection time differences, as discussed in references \cite{legero:2004, legero:2006}. 

\begin{figure}
\includegraphics[width=1.0\columnwidth,keepaspectratio]{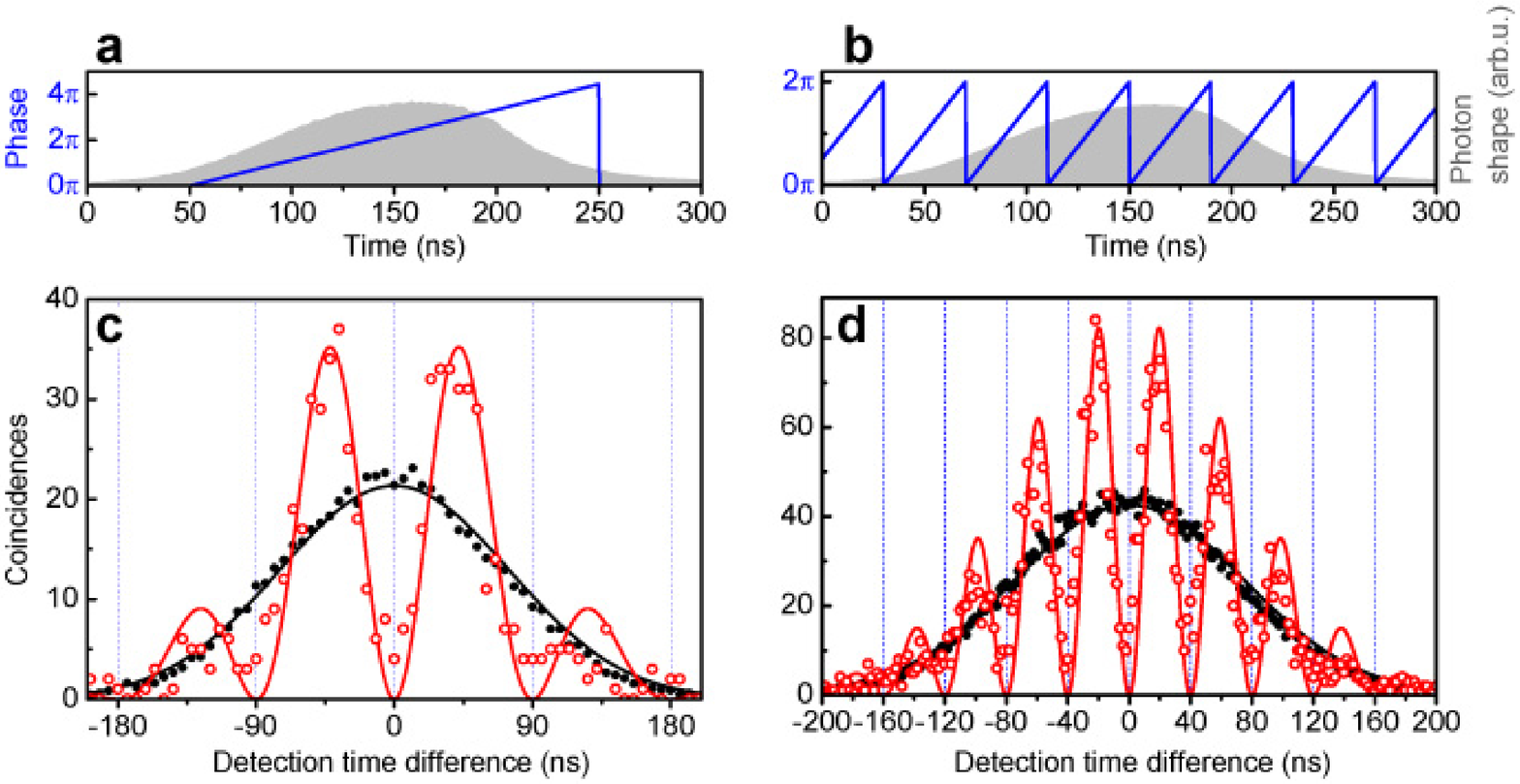}
\caption{
\textbf{Linear phase ramp.}
A linear phase ramp produces a frequency shift of the single photons, verified via a quantum beat with a second, unmodulated photon.  
\textbf{a}, A frequency shift of 11~MHz is applied via a linear phase ramp of $4.4\pi$ over the central 200~ns of the generated photon wave packets (gray shaded curve).  
\textbf{b}, A frequency shift of 25~MHz is applied by a series of 40~ns long $2\pi$ linear phase ramps.
\textbf{c,d}, Open red dots show the measured coincidences for interfering photons separated in frequency by 11~MHz (\textbf{c}) and 25~MHz (\textbf{d}).  Closed black dots show the reference data for non-interfering photons.  The black solid lines show the Gaussian shapes for non-interfering photons, calculated from the width of the photon wave packets and using only the amplitude as a fit parameter.  The red solid lines are the predicted curves for interfering, frequency-shifted photons, calculated directly from twice the amplitude of the result for non-interfering photons modulated by equation (\ref{eq:TheFormula}) with the appropriate frequency.  
All photons detected within the 300~ns windows shown in (\textbf{a}) and (\textbf{b}) are included in the data.
}
\label{fig:quantumbeat}
\end{figure}

In addition to discrete phase shifts, we also investigate the general case of continuous phase shaping within a photon wave packet.  As an example, applying a linear phase ramp to the photon traveling through the EOM results in a frequency shift, where the magnitude of the shift is controlled via the slope of the ramp.  For two photons with a frequency difference $\Delta\nu$ detected at a time difference $\tau$, the relative phase shift is $\Delta_\phi(\tau)=2\pi\Delta\nu \tau$.  Hence, the probability of photon coalescence oscillates sinusoidally at the difference frequency as a function of the detection time difference \cite{legero:2004, legero:2006, metz:2008}.

We demonstrate this phase shaping with two different phase ramp experiments.  In the first experiment, we apply a single phase ramp from 0 to $4.4\pi$ over the central 200~ns of the photon in the EOM fibre path, resulting in a frequency shift of 11~MHz (Fig.~\ref{fig:quantumbeat}a).  In the second experiment, making use of the periodicity of equation (\ref{eq:TheFormula}), a frequency shift of 25~MHz is accomplished via a saw-tooth function where each step of 40~ns introduces a phase increase of exactly $2\pi$ (Fig.~\ref{fig:quantumbeat}b).  While the finite fall time of the saw-tooth function may potentially lead to phase errors, this technique allows for arbitrarily high frequency shifts by requiring an absolute phase range of only $2\pi$ on the EOM.  In both experiments, the measured signal for interfering photons shows the expected oscillations with high visibility (Fig.~\ref{fig:quantumbeat}c-d).  

The presented phase modulation scheme not only reveals fundamental aspects of two-photon interference, but also has various potential applications.  For instance, shaping the phase of the reference photon would allow one to fully determine the phase properties of an unknown photon using this technique.  Similarly, this method to measure the phase change of single photons can be utilised for quantum key distribution protocols \cite{tittel:2000, inoue:2002}, in particular for time-bin entanglement \cite{marcikic:2002}, with the advantage of not requiring interferometric stability.  Finally, as any phase shape can be created or even compensated, this technique may be utilised for quantum computation protocols with linear optics \cite{knill:2001, wang:2006}.

\section{Methods}
\textbf{Cavity QED based single photon source.}
Individual $^{87}$Rb atoms are probabilistically transferred from a magneto-optical trap into a high-finesse optical cavity ($F\approx6\times10^4$) via a horizontally aligned dipole trap guiding beam at 1064~nm oriented orthogonal to the cavity axis \cite{bochmann:2008}.  The transit time of an atom through the cavity mode is $\approx140~\mu$s, and the probability to have one atom in the cavity at a given time is about 3\%.

To generate single photons, the intracavity atom is optically pumped to the $^2$S$_{1/2}~F=2$ hyperfine ground state via a 800~ns long laser pulse resonant with the $^2$S$_{1/2}~F=1\leftrightarrow$~$^2$P$_{3/2}~F'=2$ transition.  Next, a laser pulse resonant with the $F=2\leftrightarrow F'=1$ transition together with the cavity resonant with the $F=1\leftrightarrow F'=1$ transition drives a vacuum-stimulated Raman adiabatic passage \cite{hennrich:2000, hijlkema:2007} resulting in a single photon in the cavity mode and the atom in the $F=1$ state.  The cavity is optically asymmetric (mirror transmissions of 2~ppm and 101~ppm) such that the photon is preferentially emitted through the higher transmission mirror into the spatial output mode defined by the cavity.  The amplitude profile of the emitted photon is controlled by the temporal shape of the $F=2\leftrightarrow F'=1$ excitation laser pulse.  With an atom present in the cavity, the probability of generating a single photon in a given trial is $\approx7\%$, and the overall transmission and detection probability of this photon is also $\approx7\%$.  The insertion loss of our particular EOM is about 5~dB.

\textbf{Non-interfering photons.} 
In this manuscript, the term non-interfering photons refers to photon pairs arriving at the NPBS with no temporal overlap or with orthogonal polarisations.  Such photons exit the NPBS independently and are therefore equally likely to be detected in the same or opposite output ports.  For our reference curves, we use the average number of coincidences in the neighbouring peaks ($2\leq\vert n\vert\leq 6$) in the second-order correlation function.  With equal intensities at both input ports of the NPBS (measured to be accurate within $\pm5\%$), these coincidences occur twice as often as coincidences within the $n=0$ peak and therefore are divided by 2 for comparison \cite{maunz:2007}.  For $\vert n\vert=1$, the number of correlations is 75\% that of the outer peaks due to the probabilistic coupling of photons into either the long fibre or the EOM fibre by the PBS.

\section{Acknowledgments}
The authors thank S. Ritter for useful discussions on the manuscript.  This work was partially supported by the Deutsche Forschungsgemeinschaft (Research Unit 635, Cluster of Excellence MAP) and the European Union (IST project SCALA).  D. L. M. acknowledges support from the Alexander von Humboldt Foundation.

\textbf{Competing interests statement:} The authors declare that they have no competing financial interests.

\end{document}